\documentclass{article}
\usepackage{latexsym}
\usepackage{amssymb}
\usepackage{verbatim}
\usepackage[dvips]{graphics}

\newcommand{\mre}[0]{\mathrm{e}}
\newcommand{\mri}[0]{\mathrm{i}}
\newcommand{\mrf}[0]{\mathrm{f}}
\newcommand{\mcM}[0]{\mathcal{M}}
\newcommand{\tsum}[1]{{\textstyle \sum #1}}

\title{Beable trajectories for revealing quantum control mechanisms}

\author{ Eric Dennis, Herschel Rabitz\\
	Department of Chemistry\\
	Princeton University\\
	Princeton, NJ 08544
}

\begin{document}

\maketitle

\begin{abstract} 

The dynamics induced while controlling quantum systems by optimally shaped
laser pulses have often been difficult to understand in detail. A method
is presented for quantifying the importance of specific sequences of
quantum transitions involved in the control process. The method is based
on a ``beable'' formulation of quantum mechanics due to John Bell that
rigorously maps the quantum evolution onto an ensemble of stochastic
trajectories over a classical state space. Detailed mechanism
identification is illustrated with a model 7-level system. A general
procedure is presented to extract mechanism information directly from
closed-loop control experiments. Application to simulated experimental
data for the model system proves robust with up to 25\% noise.

\end{abstract}

\section{Introduction}

Advances in pulse shaping for ultrafast lasers, fast detection techniques,
and their integration via closed-loop algorithms have made it possible to
control the dynamics of a variety of quantum systems in the laboratory.
Excitation may be either in the strong or weak field regime, with the goal
of obtaining some desired final state. Success in achieving that goal is
gauged by a detected signal (\emph{e.g.}, the mass spectrum in the case of
selective molecular fragmentation), and this information is fed back into
a learning algorithm \cite{control}, which alters the laser pulse shape
for the next round of experiments. High duty cycles of $\sim 0.1$ seconds
or less per control experiment make it possible to iterate this process
many times and perform efficient experimental searches over a control
parameter space defining the laser pulse shape.

As an example of this process, experiments have employed closed-loop
methods for selective fragmentation and ionization of organic
\cite{organic} and organometallic \cite{metal} \cite{Na2K} compounds, as
well as for enhancing optical response in solid-state and other chemical
systems \cite{AlGaAs} \cite{Xray} \cite{dye} \cite{CH3OH}. Yields of
targeted species are typically enhanced considerably over those obtained
by non-optimized methods. It is found that the optimal pulse shapes
achieving these enhancements can be quite complicated, and understanding
their physical significance has proven difficult. The same general
observations also apply to the many optimal control design simulations
carried out in recent years \cite{Dahleh} \cite{Kosloff} \cite{Zhu}
\cite{Rice}.

The present paper will address the identification of control mechanisms in
theoretical calculations as well as for direct application in the
laboratory. In \S\ref{beables} we will first describe John Bell's beable
model for finite dimensional Hilbert spaces, in order to obtain a precise
(but non-unique) definition of ``mechanism'' for quantum systems in terms
of trajectories over their associated classical state spaces. For
instance, in molecular systems a trajectory would take the form of a
sequence of transitions that starts with a given initial molecular
configuration and switches to another configuration at a distinct time
$t_1$, and then to another at $t_2$, \emph{etc.}---to be contrasted with a
continuously changing superposition of many such configurations.

The means to numerically implement this mechanism concept is presented in
\S\ref{simulating}. An application to the problem of population transfer
for a model 7-level system is given in \S\ref{model}, which illustrates
the usefulness of mechanism information in understanding control
processes. We then show how the beable approach leads to the laboratory
working relations (\ref{jminlogM}) and (\ref{jlogM}), which make it
possible to identify some basic aspects of control mechanisms directly
from experimental data. We illustrate this process in \S\ref{simulated} on
simulated experimental data for the model 7-level system. The overall
laboratory algorithm for extracting control mechanism information is
condensed into a general-purpose procedure in \S\ref{summary}.

\section{Beables and quantum theory}\label{beables}

Consider a control problem posed in terms of the quantum evolution
\begin{equation}\label{schrod} 
\mri \hbar \frac{d}{dt}|\psi(t)\rangle = H(t)|\psi(t)\rangle 
\end{equation} 
over a finite dimensional Hilbert space with basis $|n\rangle$ where
$n=0,1,2,\ldots$. Here $H(t)=H_0-\mu E(t)$ incorporates the effect of the
control field $E(t)$, and we can explicitly follow the evolution of
$|\psi\rangle$ into a desired final state $|\psi(t_\mathrm{f})\rangle$.

This paper is concerned with the question: what is the importance of a
given sequence $n_1 \rightarrow n_2 \rightarrow \cdots$ of actual
transitions---or, more specifically, of a given trajectory defined as a
continuous function $n(t)$ of time---in achieving the desired final state
$|\psi(t_\mathrm{f}) \rangle$? In other words, it is clear that the system
\emph{is} being driven into a desired state, but can we find a physical
picture of \emph{how} this is being accomplished?

A conventional answer to the question raised above, essentially that given
by Bohr on first seeing Feynman's path integral, is to reject the question
as ill-posed because quantum mechanics is said to forbid consideration of
precisely defined trajectories over the classical state space $\{n\}$.
Nevertheless, it is well established that there exist dynamical models
generating an ensemble of trajectories $n(t)$ whose statistical properties
exactly match those associated with $|\psi(t)\rangle$ at each $t$. In the
case of a continuous state space, the first such model was that of de
Broglie, later rediscovered and completed by Bohm \cite{Bohm}. They
reintroduce classical-like particle trajectories into quantum theory by
taking the probability current $\mathbf{J}[\psi]$ to describe a
statistical ensemble of real particles. So,
\begin{equation}\label{v}  
\mathbf{v} = \frac{\mathbf{J}}{|\psi|^2}
= \mathrm{Re} \left\{
	-\mri \frac{\hbar}{m} 
	\frac{\nabla\psi(\mathbf{x},t)}{\psi(\mathbf{x},t)}
\right\}
\end{equation}
gives the velocity of a particle with mass $m$ and position $\mathbf{x}$
at time $t$, in de Broglie-Bohm (dBB) theory. The physical particle is
taken to exist independently of, but also to have its motion determined
by, the wavefunction $\psi$. The time evolution of $\psi$ itself is just
given by the Schrodinger equation.

Bohm developed a full account of how ensembles of such classical-like
particles could reproduce the predictions of quantum mechanics. A basic
issue is to compare $\psi(\mathbf{x},t)$ with the statistical distribution
$P(\mathbf{x},t)$ describing an ensemble of particles evolving by
(\ref{v}). One can show that if the initial distribution of particles
satisfies $P(\mathbf{x},0) = |\psi(\mathbf{x},0)|^2$, then
$P(\mathbf{x},t) = |\psi(\mathbf{x},t)|^2$ will hold for all $t>0$. That
is, if the ensemble is initially in the ``quantum equilibrium''
distribution given by $|\psi(\mathbf{x},0)|^2$, the dynamics---(\ref{v})
for the particles, and the Schrodinger equation for $\psi$---will preserve
this equilibrium, consistent with the predictions of standard quantum
theory. The result is easily generalized to arbitrary interacting
$N$-particle systems by taking $\mathbf{x}$ as a point in the $3N$
dimensional configuration space.

In dBB theory the position representation has a special status. While one
may still regard $\psi$ as a basis-independent object, the particle
dynamics is given by (\ref{v}) specifically in terms of $\langle
\mathbf{x} | \psi \rangle$ rather than $\langle \mathbf{p} | \psi \rangle$
or some other representation. But, it is easy to formulate analogs of dBB
theory in different bases. For instance, one might choose the momentum
values $\mathbf{p}$ as the beables\footnote{Bell used the term ``beables''
rather than the misnomer ``hidden variables'' to distinguish them from
conventional observables.} of the theory, and the dBB trajectories
$\mathbf{x}(t)$ would be replaced by momentum space trajectories
$\mathbf{p}(t)$.

In the context of a finite dimensional Hilbert space with basis
$|n\rangle$, the beables can be taken as the sites $n$ of the classical
state space $\{n\}$ analogous to $\{\mathbf{x}\}$ or $\{\mathbf{p}\}$.
Some law analogous to (\ref{v}) must be given to generate beable
trajectories $n(t)$ over the state space. Such trajectories would provide
a physical picture of the quantum transitions induced by a control field
$E(t)$.

In one such theory due to John Bell \cite{Bell}, trajectories arise from
beables stochastically jumping between sites connected by non-zero
Hamiltonian matrix elements. To define this Broglie-Bohm-Bell (BBB)
theory, the probability for a beable to jump from site $m$ to a distinct
site $n$, sometime in the interval $(t,t+\epsilon)$, is taken as

\begin{equation}\label{T}
T_{nm}(t)\epsilon = \left\{
	\begin{array}{ll}
		2\,\mathrm{Re}\{z_{nm}(t)\}\epsilon & 
			\mbox{if Re$\{z_{nm}(t)\} > 0$} \\
		0 & \mbox{if Re$\{z_{nm}(t)\} \le 0$}
	\end{array}
\right.
\end{equation}
where
\begin{equation}\label{z}
z_{nm}(t) = -\frac{\mri}{\hbar} H_{nm}
\frac{\psi_n(t)^\ast}{\psi_m(t)^\ast}
\end{equation}
and $\psi_n = \langle n|\psi\rangle$, \emph{etc.} To ensure normalization,
the probability for a beable to stay at $m$ is thus given by
$1-\sum_n^\prime T_{nm}(t)\epsilon$, where the primed sum excludes the
diagonal term $n=m$. From (\ref{z}), we find
\begin{equation}\label{Rez} \mathrm{Re}\{z_{nm}\} = -\mathrm{Re}\{z_{mn}\}
	\frac{|\psi_n|^2}{|\psi_m|^2}
\end{equation}
which implies through (\ref{T}) that \emph{either} $T_{nm}(t) = 0$
\emph{or} $T_{mn}(t) = 0$ at any particular time $t$. Together with
(\ref{schrod}) this gives
\begin{equation}\label{modpsi}
\frac{d}{dt} |\psi_n|^2 = \sum_m 2\,\mbox{Re}\{z_{nm}|\psi_m|^2\}
= \sum_m (T_{nm}|\psi_m|^2 - T_{mn}|\psi_n|^2)
\end{equation}
as a type of master equation. Note that the $T_{nm}$ term contributes when
Re$\{z_{nm}\}>0$, and the $T_{mn}$ term contributes when Re$\{z_{nm}\}<0$.

Now consider the probability distribution $P_n(t)$ of beables in state
space generated by the jump rule (\ref{T}). Accounting for the influx and
outflux of beables at site $n$, we see $P_n(t)$ satisfies
\begin{equation}\label{dfp}
\frac{d}{dt}P_n = \sum_m (T_{nm}P_m - T_{mn}P_n)
\end{equation}
which is formally identical to (\ref{modpsi}). Thus, provided $P_n(0) =
|\psi_n(0)|^2$, we are guaranteed $P_n(t) = |\psi_n(t)|^2$ for all $t>0$,
which expresses the equivalence of BBB theory and ordinary quantum
mechanics in terms of statistical predictions. 

The answer to the initial question regarding the importance of a given
trajectory in achieving the desired state $|\psi(t_\mathrm{f})\rangle$ is
now very simple. The importance may be taken as just the probability of
realizing that trajectory with the jump rule (\ref{T}). We can express the
final state population in terms of these path probabilities via the
integral (path sum) version of (\ref{dfp}):
\begin{eqnarray}
\label{pathsum}
P_{n_\mathrm{f}}(t_\mathrm{f}) &=& 
	\sum_\mathcal{P} P_{n_0}(0) \, \mathrm{Prob}(\mathcal{P})\\
\nonumber
\mathrm{Prob}(\mathcal{P}) &=&
	\prod_{p \in J} \epsilon T_{n_{p+1}n_p}
        \prod_{p \notin J} \left(
		1-\epsilon\tsum{_n^\prime} T_{n n_p}
	\right)
\end{eqnarray}
where $\mathrm{Prob}(\mathcal{P})$ is the probability of realizing the
path $\mathcal{P} = (n_0,n_1,\ldots)$ under (\ref{T}), and $n_p$ gives the
beable configuration at $t=p\epsilon$. The first sum is taken over all
such paths ending on $n_\mathrm{f}$ at $t=t_\mathrm{f}$, and
$J[\mathcal{P}] = \{p\; |\; n_{p+1} \ne n_p\}$ defines the jump set.

The above argument for the equivalence of BBB theory and ordinary quantum
mechanics ensures that the path probabilities $\mathrm{Prob}(\mathcal{P})$
are consistent with the quantum distribution $|\psi_n(t)|^2$ governing
observables. But, it should be noted that BBB theory is not unique in this
regard. The rule (\ref{T}) may be altered in non-trivial ways while
preserving the master equation (\ref{modpsi}) \cite{Guido}. The definition
(\ref{T}) might even be changed in ways that do not preserve
(\ref{modpsi}), if one is willing to relinquish a strict probability
interpretation for the trajectories \cite{Mitra1}.

In general, there are many different ways to assign probabilities to
trajectories that all result in the same time-dependent occupation
probabilities $P_n(t)$. The predictions of quantum mechanics, therefore,
cannot select a single assignment. This non-uniqueness at the root of
quantum mechanism identification can be dealt with only by reference to
the simplicity and explanatory power of a given mechanism definition.
Below we adopt the definition (\ref{T}).

\section{Simulating beables in quantum control}\label{simulating}

An ultimate goal is to obtain dynamical mechanism information directly
from experimental data associated with the closed-loop control field
optimization, without pre-existing knowledge of the system Hamiltonian or
wavefunction. Methods employing BBB theory for this purpose are presented
in \S\ref{mechanism}, but first we shall study control mechanisms for a
model system whose Hilbert space and quantum evolution are given
explicitly in numerical simulations.

Consider a quantum-optical system with level energies $\hbar \omega_n$ and
dipole moments $\mu_{nm}$.  Applying an external laser field $E(t)$, the
Hamiltonian in the interaction picture \cite{intpic} is
\begin{equation}\label{HI}
H_I = E(t) \sum_{n\,m} \mu_{nm} \mre^{\mri \omega_{nm} t}
					|n\rangle \langle m|
\end{equation}
where $\omega_{nm} \equiv \omega_n - \omega_m$. We will drop the subscript
$I$ from now on. $E(t)$ is assumed to be given by an independent
optimization algorithm designed to, for example, maximally transfer
population from $|n_\mri\rangle$ to $|n_\mathrm{f}\rangle$.

A simple second-order Schrodinger propagator was used to solve
(\ref{schrod}) in the interaction picture, relying on a factorization of
the evolution operator as
\begin{equation}\label{evol}
\mathcal{T}\left\{\mre^{-\frac{\mri}{\hbar}\int_0^t H(s)ds}\right\} = 
\prod_{p=0}^{N-1} \mathcal{T}\left\{
	\mre^{-\frac{\mri}{\hbar}\int_{t_p}^{t_{p+1}} H(s)ds}
\right\}
\end{equation}
where $t_p = p\epsilon \equiv pt_\mathrm{f}/N$ and $\mathcal{T}$ is the
time-ordering symbol. Choosing a time step $\epsilon \ll \hbar/\mu E$, we
can approximate (\ref{evol}) by dropping the $\mathcal{T}$ operations on
the right hand side and computing the integrals directly. In doing this an
error is accrued per time step given by the Baker-Hausdorf identity
$\mre^{A+B} = \mre^A\mre^B\mre^{-\frac{1}{2}[A,B]+\cdots}$ as
\begin{equation}\label{com}
\frac{1}{\hbar^2} \int_{t_p}^{t_{p+1}} \int_{t_p}^{t_{p+1}}
	 [H(r),H(s)]\,dr\,ds \; \sim \;
\left(\frac{\mu E}{\hbar}\right)^2 \epsilon^3 \omega
\, .
\end{equation}
The right hand estimate is obtained by expanding $H(r)$ to first order
about $r=s$ and noticing that the $E^\prime(s)$ term in $H^\prime(s)$
commutes with $H(s)$. The error (\ref{com}) would generally dominate third
order terms like $(\mu E \epsilon/\hbar)^3$.

If the control field is given as $E(t) = \mathrm{Re}\{\sum_i
\alpha_i E_i(t)\}$,
where
\[
E_i(t) = A(t) \mre^{\mri(\phi(t) + \omega_i^\mathrm{c}t)} 
\]
with $A(t)$ and $\phi(t)$ possibly adiabatic, we can evaluate $\int
H(s)ds$ by writing
\begin{equation}\label{intA}
\int_{t_p}^{t_{p+1}} \mu E_i(s) \mre^{\mri \omega s} ds
\; \approx \;
\frac{ \mu A(t_p) \mre^{\mri\phi(t_p)} }
	{ \mri ( \omega + \omega_i^\mathrm{c} ) }
	\left( 
	\mre^{ \mri (\omega + \omega_i^\mathrm{c}) t_{p+1} } 
	- \mre^{ \mri (\omega + \omega_i^\mathrm{c}) t_p } 
	\right)
\; .
\end{equation}
(Simply writing $\int H(s)ds \approx \epsilon H(t_p)$ is not appropriate
because we do not want to exclude weak field excitation, \emph{i.e.}\ $\mu
E \ll \hbar \omega$, so that $\omega \epsilon \sim 1$ may hold.) Thus in
the adiabatic case $|\psi(t)\rangle$ can be propagated in steps determined
by $A(t)$ and $\phi(t)$ rather than the phase factors $\mre^{\mri\omega
t}$.

Consider the evolution of beable trajectories according to (\ref{T}),
which appears to require a time step small enough that each part of $H$,
including the $\mre^{\mri\omega t}$ terms, not vary much over the step.
Nevertheless, the total probability of jumping from $m$ to $n$ over
$(t_p,t_{p+1})$ is given by the integral $\int T_{nm}(s)ds$ over that
range with $\sim (\mu E\epsilon/\hbar)^2$ corrections. Thus we can take an
effective jump probability for the interval $(t_p,t_{p+1})$ as given by
(\ref{T}) with
\begin{equation}\label{zint}
z_{nm}(t_p) \; \approx \;
-\frac{\psi_n(t_p)^\ast}{\psi_m(t_p)^\ast}
	\frac{\mri}{\hbar \epsilon} \int_{t_p}^{t_{p+1}} H_{nm}(s) ds
\end{equation}
evaluated using (\ref{intA}). If $\omega\epsilon \ll 1$ does not hold,
care must be taken to extend the integration in (\ref{zint}) only over $t
\in (t_p,t_{p+1})$ for which $\mathrm{Re}\{z_{nm}(t)\}>0$, leading to
additional boundary terms in the phase difference part of (\ref{intA}).
Moving the $\psi^\ast$ ratio outside the integral in (\ref{zint}) produces
an error per time step of order
\[
\frac{\epsilon^2 H}{\hbar} 
\frac{\partial \psi}{\partial t} 
\; \sim \;
\left(\frac{\mu E \epsilon}{\hbar}\right)^2
\]
which is again comparable to (\ref{com}). Therefore beable trajectories
may be propagated in steps determined by $A(t)$ and $\phi(t)$,
\emph{i.e.}\ in sink with the Schrodinger propagator.

\section{A model 7-level system: mechanism analysis of an optimal control
design}\label{model}

\begin{figure}
\centering
\scalebox{.5}{\includegraphics{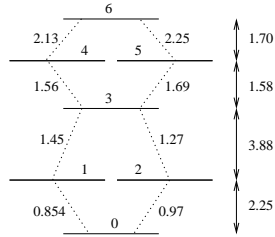}}
\caption{\small The model 7-level system $|n\rangle$ with $n
=0,1,\ldots,6$. The transition frequencies $\omega_{nm}$ in units of
fs$^{-1}$ are shown on the right, and non-zero dipole matrix elements
$\mu_{nm}$ in units of $10^{-30}$ C$\cdot$m are indicated by dotted lines.} 
\label{7levels}
\end{figure}

\begin{figure}
\centering
\scalebox{.75}{\includegraphics{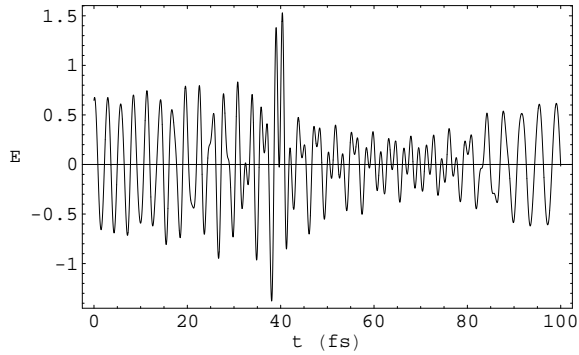}}
\caption{\small 
Electric field $E(t)$ in $V/\mathrm{\AA}$ obtained from an
optimization algorithm for population transfer from $|0\rangle$ to
$|6\rangle$ \cite{Mitra2}.
}
\label{Efield}
\end{figure}

\begin{figure}
\centering
\scalebox{.75}{\includegraphics{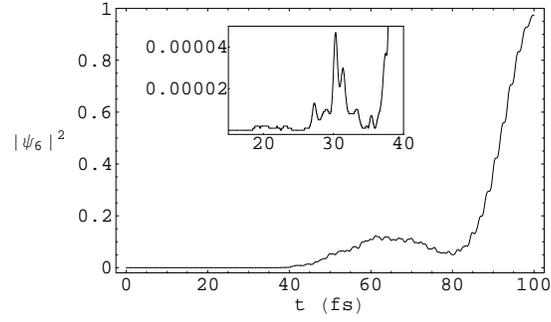}}
\caption{\small 
Population $|\psi_6(t)|^2$ as a function of time. Detail for
small $t$ is shown in the inset.
}
\label{psi6}
\end{figure}

The beable trajectory methodology for identification of control mechanisms
will be illustrated with a 7-level system where $\omega_n$ and $\mu_{nm}$
are given in Fig.\ \ref{7levels}. The (non-adiabatic) control field $E(t)$
shown in Fig.\ \ref{Efield} is obtained from a steepest descents algorithm
over the space of field histories \cite{Mitra2}. It is optimized to
transfer population from the ground state $|0\rangle$ to the highest
excited state $|6\rangle$. By $t = 100$ fs, the transfer is found to be
completed with approximately 97\% efficiency (see Fig.\ \ref{psi6}).

\begin{figure}
\centering
\scalebox{.75}{\includegraphics{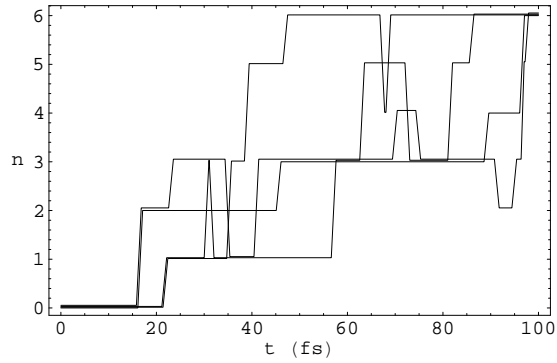}}
\caption{\small 
One each of the 4, 6, 8, and 10-jump trajectories generated by the jump
rule (\ref{T}) are shown with their sites $n$ plotted against time. For
viewing purposes, we have displaced them a small amount vertically from
each other and tilted the jump lines slightly away from the vertical.
} 
\label{4traj}
\end{figure}

Together with the second-order Schrodinger propagator, using time step
$\epsilon = .025$ fs, an ensemble of $N_{\mathrm{traj}} = 10^5$ beable
trajectories is evolved, all starting in the ground state $n=0$ at $t=0$.
At each time step, a given beable at site $m$ is randomly made either to
jump to a neighboring site $n \ne m$ according to the probabilities
$T_{nm}\epsilon$ given by (\ref{T}) with (\ref{zint}), or else stay at
$m$. Four sample trajectories are shown in Fig.\ \ref{4traj}. As a check,
one can count the number of beables residing on each site $n$ at time $t$
to estimate the occupation probabilities $P_n(t)$ and verify that they
match the quantum prescriptions $|\psi_n(t)|^2$. The finite-ensemble
deviations are observed to be consistent with a
$(N_{\mathrm{traj}})^{-1/2}$ convergence law.

About 60\% of the trajectories generated are found to involve four jumps,
and of these the trajectories passing through sites $n=2,5$ are noticeably
more probable than those passing through $n=1,4$. 6-jump trajectories
comprise about 30\% of the ensemble. And it becomes increasingly less
likely to find trajectories with more and more jumps. The largest number
of jumps observed in a single trajectory was 14. Three such trajectories
occurred out of the ensemble total $10^5$.

A natural expectation is that the optimal field $E(t)$ would concentrate
on the higher probability trajectories and not waste much effort on
guiding highly improbable trajectories, such as the 14-jumpers, to the
target state $n=6$, as the latter have essentially no impact on the
control objective (final population of the target state). Interestingly,
though, the vast majority of even the lowest probability trajectories are
still guided to $n=6$. Apparently, the optimal field is able to coordinate
its effect on low probability trajectories with that on other trajectories
at no real detriment to the latter. We shall come back to this point
later.

One way to conveniently categorize the large set of trajectories, each
expressible as a sequence of time-labeled jumps $(t_1,n_1) \rightarrow
(t_2,n_2) \rightarrow \cdots$, is to drop the time labels, leaving only
the ``pathway'' $n_1 \rightarrow n_2 \rightarrow \cdots$. The importance
of a given pathway is then computed as the frequency of trajectories
associated with that pathway. Table \ref{tab} lists some important and/or
interesting pathways and their probabilities.  

\begin{figure}
\centering
\begin{tabular}{cc}
\scalebox{.6}{\includegraphics{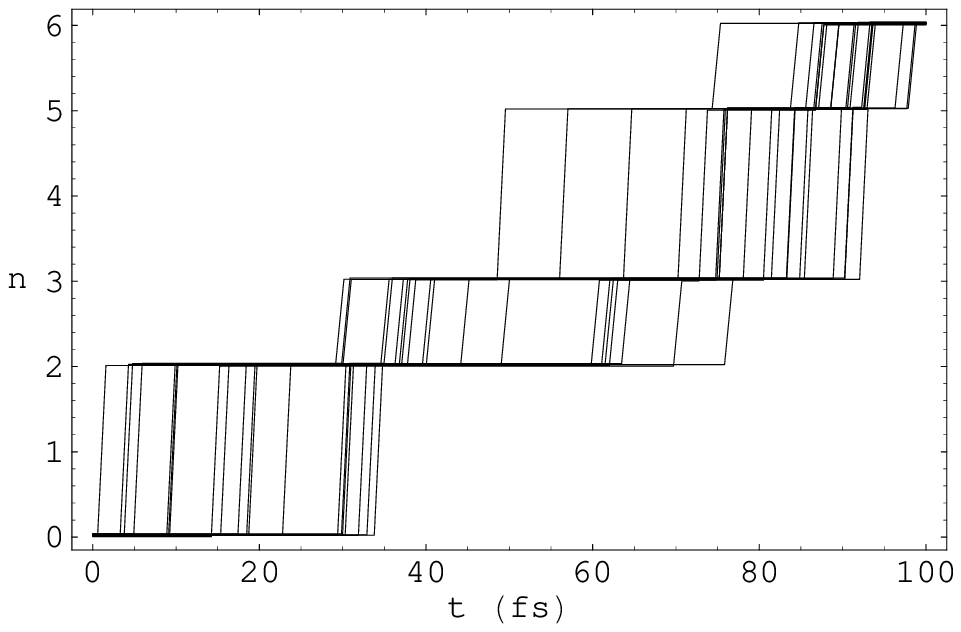}} &
\scalebox{.6}{\includegraphics{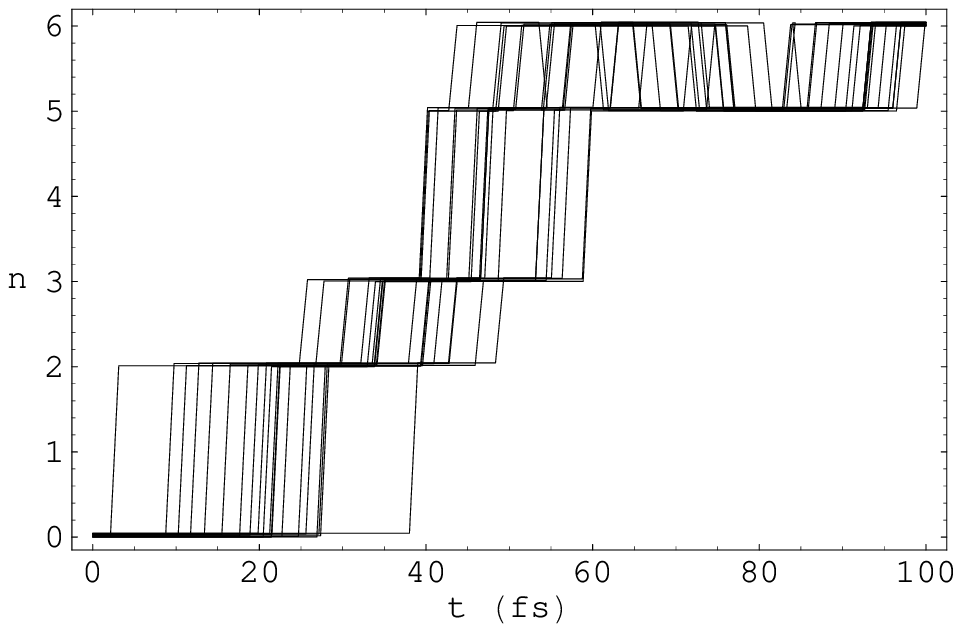}} \\
\end{tabular}
\caption{\small 
A sample of 20 trajectories each from the pathways 0~2~3~5~6 and
0~2~3~5~6~5~6.
} 
\label{46jumpers}
\end{figure}

Fig.\ \ref{46jumpers} shows some typical trajectories associated with the
first and fifth pathways listed in Table \ref{tab}---involving 4 and 6
jumps respectively. $E(t)$ guides the 4-jumpers upward in energy, and they
begin to arrive at $n=6$ around $t = 80$ fs, early enough that stragglers
can catch up but too late for the over-achievers of the group to head off
elsewhere. This corresponds to the onset of heavy growth for
$|\psi_6(t)|^2$ around $t = 80$ fs (see Fig.\ \ref{psi6}). The 6-jumpers
first reach $n=6$ around $t = 50$ fs, but almost all fall back to $n=5$ by
$t=80$ fs, reuniting with the 4-jumpers just as they begin to jump up to
$n=6$. These 6-jumpers, along with other high-order contributions, thus
explain the small surge in $|\psi_6(t)|^2$ between 50 and 80 fs. Another
much smaller surge around $t = 30$ fs and one still smaller around $t =
20$ fs (see inset of Fig.\ \ref{psi6}) are attributable to 8-th and higher
order trajectories ``ringing'' back and forth on $5 \leftrightarrow 6$.

For $t \in (70\mbox{ fs},80\mbox{ fs})$, many of the 6-jumpers are at
$n=6$ and need to be de-excited on the $6 \rightarrow 5$ transition before
they can jump back up to $n=6$. Simultaneously, many of the 4-jumpers are
at $n=5$ and should not be prematurely excited on $5 \rightarrow 6$, lest
they not remain at $n=6$ through $t=100$ fs. The optimal field thus faces
a conundrum: how to stimulate the $2 \leftrightarrow 6$ transition
preferentially for the 6-jumpers (in $n=6$) over the 4-jumpers (in $n=5$).
The means by which this feat is accomplished may be understood by
reference to the jump rule (\ref{T}). $E(t)$ induces jumps through the
explicit $H_{nm}(t)$ factor but also through the $\psi^\ast$ quotient,
which depends on $E(t)$ through $(\ref{schrod})$. In particular,
(\ref{Rez}) implies that at any one time $t$ jumps on this transition must
be either all upward or all downward. The active direction is switched
back and forth according to the sign of $\mathrm{Re}\{z_{65}(t)\}$.

\begin{figure}
\centering
\scalebox{.75}{\includegraphics{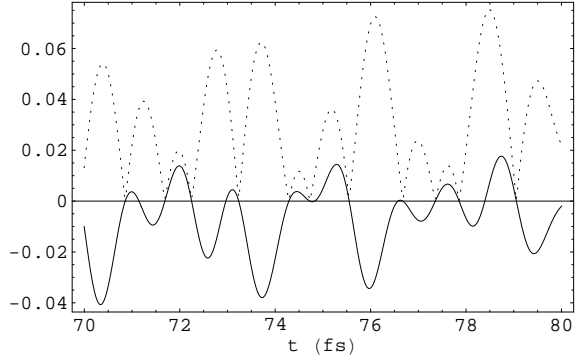}}

\caption{\small 
The optimal field modulus $|E(t)|$ (dotted line) and
$\mathrm{Re}\{z_{65}(t)\}$ (full line) in $\mbox{fs}^{-1}$ over the range
$(70\mbox{ fs}, 80\mbox{ fs})$. Their anticorrelation causes beables to be
preferentially selected for the downward transition $6 \rightarrow 5$ over
the upward transition $5 \rightarrow 6$.
} 
\label{RezEfield}
\end{figure}

Fig.\ \ref{RezEfield} plots $|E(t)|$ and $\mathrm{Re}\{z_{65}(t)\}$, which
controls the upward jump rate $T_{65}(t)$. For $t \in (70\mbox{
fs},80\mbox{ fs})$ one sees that when $|E(t)|$ is large, most often
$\mathrm{Re}\{z_{65}(t)\}$ dips below zero, disallowing any upward jumps.
The correlation coefficient between $|E(t)|$ and
$\mathrm{Re}\{z_{65}(t)\}$ in this range is $-0.4955$. On the other hand,
the correlation between $|E(t)|$ and $\mathrm{Re}\{z_{56}(t)\}$, which
controls downward jumping, is $+0.4475$ over the same range.

\begin{figure}
\centering
\scalebox{.75}{\includegraphics{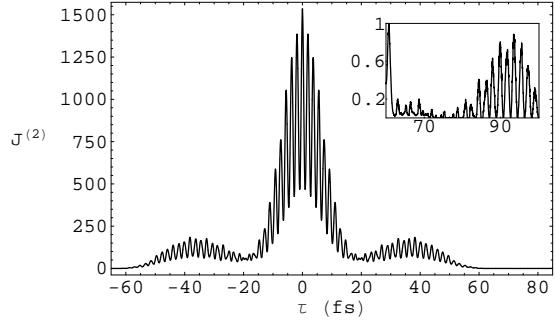}}
\caption{\small 
Jump correlation function $J_{\Omega}^{(2)}(\tau)$ associated with jumps
on $5 \rightarrow 6$, plotted against the delay time $\tau$ for the
ensemble of $10^5$ trajectories. Detail for large $\tau$ is
shown in the inset.
} 
\label{J2}
\end{figure}

Looking at the trajectories in more detail, one notices a distinct
bunching of jumps. Beables tend to jump together in narrow time bands, or
else to abstain in unison from jumping. This behavior can be gauged
by calculating the two-time jump-jump correlation function:
\[
J_{\Omega}^{(2)}(\tau) \equiv 
\frac{1}{N}\sum_{p=0}^{N-1} J_{\Omega}(t_p)J_{\Omega}(t_p + \tau)
\]
where $J_{\Omega}(t)$ is the number of jumps of type $\Omega$ occurring in
$(t,t+\epsilon)$, and $\Omega$ is a subset of the entire ensemble of
trajectories. For instance, the two-time function with $\Omega$ taken as
the set of jumps on the $5 \rightarrow 6$ transition is plotted in Fig.\
\ref{J2}. The fs time-scale oscillations correspond to the level
splittings $\omega_{nm}$ and the dominant frequency components of $E(t)$.
Enhanced correlations around $\tau=0$ correspond to the jump bunching
noticeable in the trajectories. Two side-bands around $\tau = \pm 40$~fs
are associated with 6-jump and higher order trajectories that go up, down,
and up again on $5 \leftrightarrow 6$ over the approximate time window
$(50\mbox{ fs},90\mbox{ fs})$. This conclusion can be verified by
computing two-time functions with $\Omega$ specialized to particular
pathways. Other much smaller features for $|\tau| > 60$ fs (see inset of
Fig.\ \ref{J2}) are attributable to higher order trajectories ringing on
$5 \leftrightarrow 6$.

In general, the fs oscillations characteristic of these two-time functions
show that $E(t)$ works in an essentially discrete way, turning on the flow
of beables over a given transition and then turning it off with a duty
cycle of $\approx 2$~fs. The associated bandwidth of $\approx 0.5 \mbox{
fs}^{-1}$ is small enough to discriminate between all non-degenerate
$\omega_{nm}$ except between $\omega_{35} (=\omega_{34})$ and $\omega_{56}
(=\omega_{46})$, which differ by only $0.12\mbox{ fs}^{-1}$. This
circumstance leaves effectively three distinguishable transitions. With a
total time of 100 fs, the control field $E(t)$ can potentially enact
roughly $150$ separate flow operations. The fact that trajectories with
pathway probability $\ll 1$\% are still almost always guided successfully
to $n=6$ suggests that these $\sim 150$ operations are more than necessary
to obtain the 97\% success rate achieved by the optimal control algorithm
in this simulation. It appears that the algorithm actively sweeps these
aberrant trajectories back into the mainstream so as to maximize even
their minute contribution to the control objective.

\section{Control mechanism identification in the
laboratory}\label{mechanism}

Using these beable trajectory methods to extract mechanism information
directly from closed-loop data is complicated by the fact that we cannot
assume knowledge of a time-dependent wavefunction, Hamiltonian, or
possibly even the energy level structure of the system. Frequently in the
laboratory, the only available information consists of final state
population measurements and knowledge of the control field $E(t)$.

The following analysis aims to show how a limited statistical
characterization of beable trajectories may be generated from laboratory
data associated with a given optimal control field. In particular, we will
show how to extract $j_\mathrm{min}$, the minimum number of jumps
necessary to reach the final state $n_\mathrm{f}$ from the initial state
$n_\mathrm{i}$; also $\langle j_\mathcal{P} \rangle$, the average number
of such jumps over an ensemble of beable trajectories; and possibly higher
moments $\langle (j_\mathcal{P})^k \rangle$ as well. After a general
formulation of this analysis is presented, it will be applied to simulated
experimental data in the case of the model 7-level system considered
above.

We propose to obtain mechanism information by examining the effect on the
final state population $|\psi_{n_\mrf}|^2$ of variations in the control
field \emph{away from} optimality. Consider the simplest such scheme,
wherein the amplitude of the control field is modulated by a constant
$\mcM$ independent of time:
\[
E(t) \rightarrow \tilde E(t) = \mcM E(t)
\]
giving rise to a new time-dependent solution $|\tilde \psi(t)\rangle$---in
particular, a new final state population $|\tilde \psi_{n_\mrf}
(t_\mrf)|^2$ and new path probabilities P$\widetilde{\mathrm{rob}}
(\mathcal{P})$. These quantities are obtained by taking $T_{nm}
\rightarrow \tilde T_{nm}$ in (\ref{pathsum}), which is to say using
$\tilde E(t)$ and $\tilde \psi_n(t)$ in the jump rule (\ref{T}).

To express P$\widetilde{\mathrm{rob}}(\mathcal{P})$ in terms of 
Prob$(\mathcal{P})$, we can write
\begin{eqnarray}
\prod_{p \in J} \tilde T_{n_{p+1}n_p} &=& 
\nonumber
	\mcM^{j_\mathcal{P}} \prod_{p \in J} T_{n_{p+1}n_p}
	\prod_{p \in J} \frac{\cos \tilde \phi_p}{\cos \phi_p} \; \times\\
\label{prodTx}	&& \; \;
	\prod_{p \in J} 
		\frac{|\tilde \psi_{n_{p+1}}(t_p)|}{|\tilde
		\psi_{n_p}(t_p)|} 
	\left(\prod_{p \in J} 
		\frac{|\psi_{n_{p+1}}(t_p)|}{|\psi_{n_p}(t_p)|}
	\right)^{-1}
\end{eqnarray}
where $j_\mathcal{P}$ is the number of jumps in $\mathcal{P}$ and 
\[
\tilde \phi_p \; \equiv \; \arg\left(
	-\mri H_{n_{p+1}n_p} 
	\frac{\tilde \psi_{n_{p+1}}(t_p)}{\tilde \psi_{n_p}(t_p)} 
\right) \, .
\]
To simplify (\ref{prodTx}), note that if $j_\mathcal{P}$ were very large,
then successive terms in each of the last two products would tend to
cancel, leaving only endpoint contributions. Making the reasonable
approximation that they do completely cancel yields
\begin{equation}\label{prodT}
\prod_{p \in J} \tilde T_{n_{p+1}n_p} \; \approx \; 
	\mcM^{j_\mathcal{P}} \,
	\frac{|\tilde \psi_{n_\mrf}(t_\mrf)|}{|\psi_{n_\mrf}(t_\mrf)|} \,
	\prod_{p \in J} T_{n_{p+1}n_p}
	\prod_{p \in J} \frac{\cos \tilde \phi_p}{\cos \phi_p} \, .
\end{equation}
Further, we can make the expansion
\[
-\log \prod_{p \in J} \frac{\cos \tilde \phi_p}{\cos \phi_p} 
\; = \;
a_\mathcal{P}^{(1)}(\mcM-1) + a_\mathcal{P}^{(2)}(\mcM-1)^2 + \cdots
\]
about $\mcM=1$, where the $a_\mathcal{P}^{(i)}$ depend on the path
$\mathcal{P}$ but not on $\mcM$. And similarly:
\begin{eqnarray*}
-\log \prod_{p \notin J} \left(
	1-\epsilon\tsum{_n^\prime} \tilde T_{n n_p}
\right) 
&\approx&
\epsilon \tsum{_{p \notin J}} \tsum{_n^\prime} \tilde T_{n n_p}\\
&=&
\epsilon \tsum{_{p \notin J}} \tsum{_n^\prime} T_{n n_p} +
b_\mathcal{P}^{(1)}(\mcM-1) + \cdots
\end{eqnarray*}
Combining these expansions gives a relationship between the path
probabilities P$\widetilde{\mathrm{rob}}(\mathcal{P})$ in the modulated
case to those, Prob$(\mathcal{P})$, in the unmodulated case, which are the
ones containing mechanism information regarding the actual optimal control
field $E(t)$. We can thus write the final population as
\begin{eqnarray*}
|\tilde\psi_{n_\mrf}(t_\mrf)|^2 &=&
\sum_\mathcal{P} |\psi_{n_0}(0)|^2
\, \mbox{P$\widetilde{\mathrm{rob}}(\mathcal{P})$}\\
& \approx &
\frac{|\tilde \psi_{n_\mrf}(t_\mrf)|}{|\psi_{n_\mrf}(t_\mrf)|}
\sum_\mathcal{P} |\psi_{n_0}(0)|^2 \mcM^{j_\mathcal{P}}
\mre^{-a_\mathcal{P}(\mcM-1)} \, \mathrm{Prob}(\mathcal{P})
\end{eqnarray*}
where $a_\mathcal{P} \equiv a_\mathcal{P}^{(1)} + b_\mathcal{P}^{(1)}$,
and higher order terms in the expansion have been dropped. (This
approximation is not as crude as it might seem, since for small $\mcM$
away from 1, the behavior of $|\tilde\psi_{n_\mrf}(t_\mrf)|^2$ is
dominated by the $\mcM^{j_\mathcal{P}}$ factor.) Cancelling one power of
$|\tilde \psi_{n_\mrf}(t_\mrf)|$, and recalling that the sum is taken only
over paths ending on $n=n_\mrf$ so that $|\psi_{n_\mrf}(t_\mrf)|^2 =
\sum_\mathcal{P} \mathrm{Prob}(\mathcal{P})$, we have
\begin{equation}\label{M^j}
|\tilde\psi_{n_\mrf}(t_\mrf)|
\; \approx \;
|\psi_{n_\mrf}(t_\mrf)| \left\langle 
	\mcM^{j_\mathcal{P}} \mre^{-a_\mathcal{P}(\mcM-1)}
\right\rangle
\end{equation}
where $\langle \cdots \rangle$ denotes an average over the trajectory
ensemble generated by the (unmodulated) optimal field $E(t)$. Beables in
this ensemble are taken as initially distributed at $t=0$ according to
$|\psi_{n}(0)|^2$, and only trajectories that successfully reach
$n=n_\mrf$ at $t=t_\mrf$ are counted.

Note that for $\mcM$ close enough to 0, the minimum value $j_\mathrm{min}$
taken on by $j_\mathcal{P}$ will dominate the expectation value in
(\ref{M^j}), and
\begin{equation}\label{jminlogM}
\log |\tilde\psi_{n_\mrf}(t_\mrf)| = j_\mathrm{min} \log\mcM +
\mathrm{O}(1)
\end{equation}
gives the dominant behavior independent of $a_\mathcal{P}$. If we suppose
that $a_\mathcal{P}$, where it is relevant, depends primarily on the
endpoints of $\mathcal{P}$, which are fixed, and only weakly on the rest
of the path, then $a_\mathcal{P}$ can be approximated by some
characteristic value $a$. Putting $\mcM^{j_\mathcal{P}} =
\mre^{j_\mathcal{P} \log\mcM}$ in (\ref{M^j}) and expanding in powers of
$\log\mcM$ now gives
\begin{equation}\label{jlogM}
|\tilde\psi_{n_\mrf}(t_\mrf)| 
\; \approx \;
|\psi_{n_\mrf}(t_\mrf)| \, \mre^{-a(\mcM-1)} 
\sum_{k=0}^\infty \frac{\langle (j_\mathcal{P})^k \rangle}{k!}
	(\log\mcM)^k
\end{equation}
for the final state population under a modulated field, expressed
in terms of the desired statistical properties of the trajectory ensemble 
under the optimal field itself. Here, $a$ enters as an additional
parameter that must be extracted from the data. Equations (\ref{jminlogM})
and (\ref{jlogM}) form the working relations to extract mechanism
information from laboratory data.

\section{Simulated experiments on a 7-level system}\label{simulated}

In order to extract quantities like $\langle j_\mathcal{P} \rangle$ using
the results (\ref{jminlogM}) and (\ref{jlogM})  data must be generated for
the final state population $|\tilde \psi_{n_\mrf}(t_\mrf)|^2$ at many
values of the modulation factor $\mcM$ over some range
$(\mcM_\mathrm{min},\mcM_\mathrm{max}) \sim (0,1.5)$. The desired
quantities are obtained as parameters in fitting (\ref{jminlogM}) and
(\ref{jlogM})  to the data as a function of $\mcM$.

\begin{figure}
\centering
\scalebox{.75}{\includegraphics{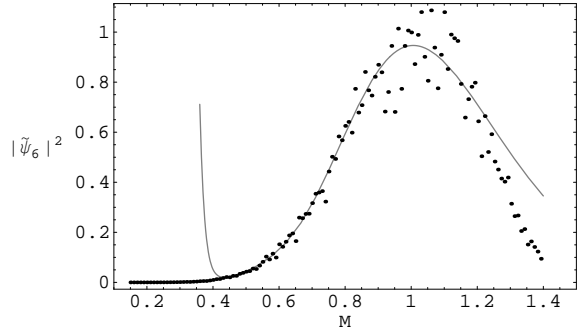}}
\caption{\small 
The best fit of (\ref{jlogM}) to the simulated $|\tilde
\psi_{n_\mrf}(t_\mrf)|^2$ data (10\% noise) as a function of $\mcM$; it
occurs over the fitting range $\mcM \in (.44,.92)$.
} 
\label{fit}
\end{figure}

One set of simulated data for the above 7-level system is shown in Fig.\
\ref{fit}; the sampling increment is $\Delta M = .01$. Noise has been
introduced by multiplying the exact $|\tilde \psi_{n_\mrf}(t_\mrf)|^2$
values by an independent Gaussian-distributed random number for each value
of $\mcM$, where the distribution is chosen to have mean 1, and various
standard deviations $\sigma$ have been sampled.

\begin{figure}
\centering
\scalebox{.75}{\includegraphics{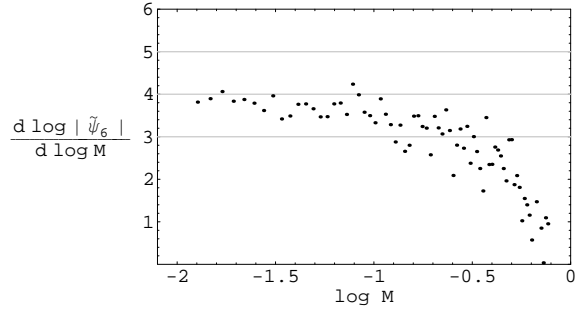}}
\caption{\small 
The derivative is calculated from simulated data with
noise level $\sigma=.1$; its limiting value as $\log\mcM \rightarrow
-\infty$ gives $j_\mathrm{min}$. 
} 
\label{jmin}
\end{figure}

We can determine $j_\mathrm{min}$ from the data using (\ref{jminlogM}),
which implies 
\begin{equation}\label{dlogpsidlogM}
j_\mathrm{min} = \lim_{\mcM \rightarrow 0} 
	\frac{d \log |\tilde \psi_{n_\mrf}(t_\mrf)|}{d\log\mcM} \, .
\end{equation}
For instance, Fig.\ \ref{jmin} plots the derivative in
(\ref{dlogpsidlogM}), calculated with finite differences from the $|\tilde
\psi_{n_\mrf}(t_\mrf)|^2$ simulated data for $\sigma=.1$, which correctly
gives $j_\mathrm{min}=4$ as the limiting value. Determination of
$j_\mathrm{min}$ proved robust to multiplicative Gaussian noise up to the
40\% level ($\sigma=.4$).

The quantity $\langle j_\mathcal{P} \rangle$ is more difficult to extract,
because while the sum in (\ref{jlogM}) converges to 0 as $\mcM \rightarrow
0$, the terms of the sum individually diverge and must cancel in a
delicate manner. Therefore truncating the sum to an upper limit
$k_\mathrm{max}$ becomes a very bad approximation near $\mcM=0$. This
unstable behavior can be controlled by carefully setting the range
$(\mcM_\mathrm{min}, \mcM_\mathrm{max})$ of data to be fitted, given a
choice of $k_\mathrm{max}$.

It is also convenient to constrain the fit by the previous determination
of $j_\mathrm{min} = 4$. We have done this by noting that if $E(t)$ is
truly optimal, then $|\tilde \psi_{n_\mrf}(t_\mrf)|$ must have a maximum
at $\mcM=1$, which implies that $a = \langle j_\mathcal{P} \rangle$. This
can be used as a weaker constraint on the auxiliary parameter $a$ by just
requiring $a > j_\mathrm{min} = 4$ in the fit without necessarily
supposing that $E(t)$ is exactly optimal. We then check that $a \approx
\langle j_\mathcal{P} \rangle$ is satisfied in the fit. Fig.\ \ref{fit}
shows one such fit where the fitting range is $\mcM \in (.44,.92)$. One
can see that the fit closely tracks the data for $\mcM$ in this range but
quickly diverges from the data just below $\mcM = .44$ (and, less
severely, above $\mcM = .92$) due to the sum-truncation instability
mentioned previously.

In order to identify appropriate ranges in general, we have searched over
all combinations such that 
\begin{equation}\label{range}
\begin{array}{c}
	\begin{array}{lcccr}
	.2 &<& \mcM_\mathrm{min} &<& .8 \\ 
	.7 &<& \mcM_\mathrm{max} &<& 1.6 \\
	\end{array}\\
\mcM_\mathrm{max}-\mcM_\mathrm{min}>10
\end{array}
\end{equation}
Mathematica's implementation of the Levenberg-Marquardt non-linear fitting
algorithm was used on simulated data for each value of $\sigma$ between 0
and .5 with a .01 increment. The best fit at each $\sigma$ was used to
determine the value of $\langle j_\mathcal{P} \rangle$ most consistent
with the simulated data at the given noise level. 

For this analysis $k_\mathrm{max}=4$ was chosen somewhat arbitrarily
to balance computational cost and precision. In practice it is likely that
the moments $\langle (j_\mathcal{P})^k \rangle$ for lower $k$ values will
be most reliably extracted from the data, especially considering the
laboratory noise. In the simulations it was found that $\langle
j_\mathcal{P} \rangle$ could be reliably extracted, but higher moments
were unstable and unreliable. For example, $\langle (j_\mathcal{P})^2
\rangle$ was frequently found to lie slightly below the corresponding fit
values for $\langle j_\mathcal{P} \rangle^2$, which is inconsistent with
the interpretation of these values as statistical moments of an underlying
random variable $j_\mathcal{P}$. Further constraints could be introduced
to attempt to stabilize the extraction of higher moments, but care is
needed so as not to overfit the data.

\begin{figure}
\centering
\scalebox{.75}{\includegraphics{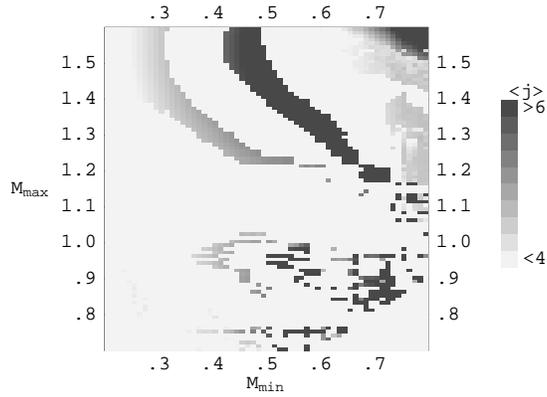}}
\caption{\small 
Fit values for $\langle j_\mathcal{P} \rangle$ over a set of different
fitting ranges $(\mcM_\mathrm{min},\mcM_\mathrm{min})$; $\sigma=.1$ here. 
}
\label{javg}
\end{figure}

\begin{figure}
\centering
\scalebox{.75}{\includegraphics{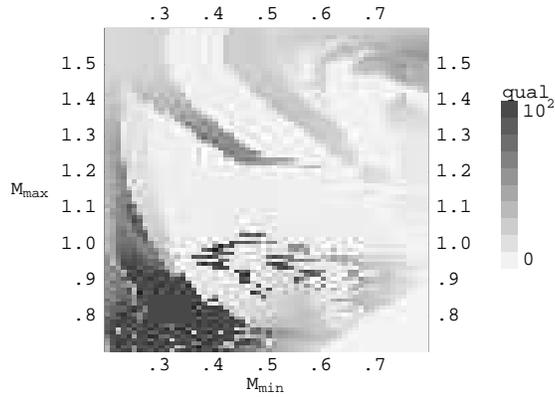}}
\caption{\small 
Fit qualities as measured by the inverse of the mean squared deviations
between the simulated data and the fit; $\sigma=.1$ here.
The highest fit quality appears at $(.44,.92)$. 
} 
\label{fq}
\end{figure}

Fig.\ \ref{javg} shows the $\langle j_\mathcal{P} \rangle$ values obtained
by fitting data with $\sigma=.1$ for each choice of $(\mcM_\mathrm{min},
\mcM_\mathrm{max})$ and Fig.\ \ref{fq} shows the corresponding quality of
each fit as measured by its mean squared deviations. In Fig.\ \ref{javg},
as well as in the corresponding plots for all other values of $\sigma$
studied, two diagonal strips emerge running above a set of smaller
islands. The surrounding white ``sea'' comprises fits that give $\langle
j_\mathcal{P} \rangle < 4$, which we know to be ruled out by the
determination of $j_\mathrm{min}$. 

A virtually identical pattern arises in the fit quality plots. The two
strips and underlying islands are seen to give much better fits than the
white sea. An additional connected region of good fits is found to extend
across the lower-left corner of Fig.\ \ref{fq}, nearly all of which are
ruled out by $j_\mathrm{min}=4$. This connected region is somewhat
pathological because much of it corresponds to fitting ranges that fail to
capture the important behavior of $|\tilde \psi_{n_\mrf}(t_\mrf)|^2$ near
$\mcM=1$, and therefore can be ignored. Then the best fits for all values
of $\sigma$ sampled are found to come from the cluster of islands at
$\mcM_\mathrm{max} \approx .95$. As $\sigma$ is increased from 0 to .5,
these islands flow from $\mcM_\mathrm{min} \approx .5$ to
$\mcM_\mathrm{min} \approx .3$, carrying with them the best fit site. Note
that the small triangular area in the lower right corner, most noticeable
in Fig.\ \ref{fq}, is a region excluded from consideration by the third
constraint in (\ref{range}).

\begin{figure}
\centering
\scalebox{.75}{\includegraphics{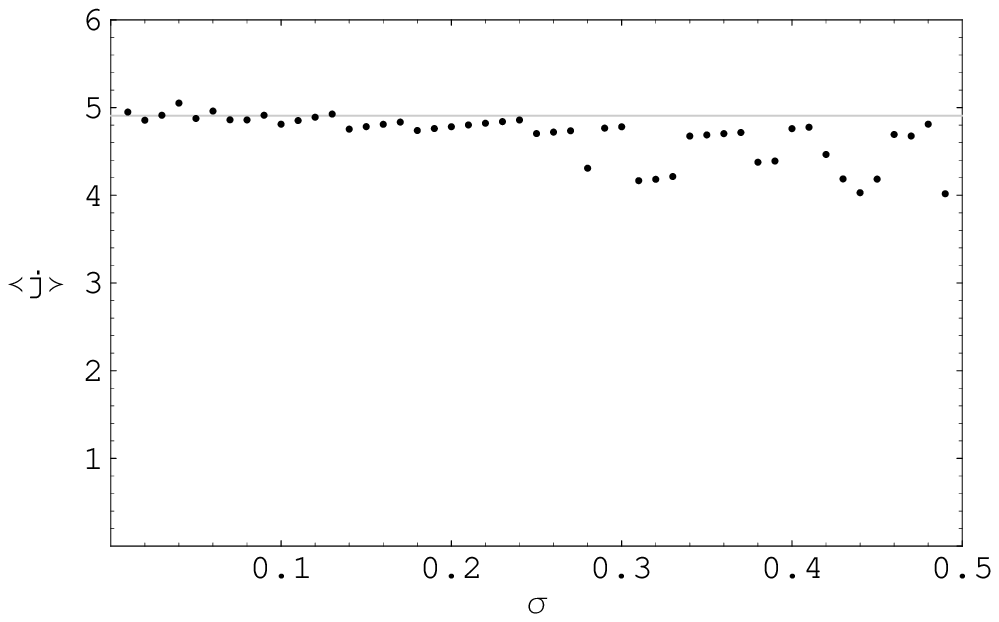}}
\caption{\small 
Best fit values for $\langle j_\mathcal{P} \rangle$ as a function of the
noise level $\sigma$, compared to the exact value $\langle j_\mathcal{P}
\rangle = 4.907$ (gray line).
} 
\label{javgsigma}
\end{figure}

The best fit values of $\langle j_\mathcal{P} \rangle$ are shown as a
function of $\sigma$ in Fig.\ \ref{javgsigma}. These values are to be
compared with the exact result $\langle j_\mathcal{P} \rangle = 4.907$
obtained from the trajectory ensemble calculations in \S\ref{model}, which
require explicit knowledge of the level structure and dipole moments
$\mu_{nm}$ of the system. The ramping behavior in Fig.\ \ref{javgsigma}
results from the sampling increment $\Delta\mcM$ of the simulated data.
Transitioning between one ramp and another corresponds to the shifting of
the best fit location by one or two units of $\Delta\mcM$.

These $\langle j_\mathcal{P} \rangle$ values are in good agreement (3\%
discrepancy) with the exact value for noise at the level of 0--25\%. It
should be noted that a qualitative change occurs in the case of no noise
($\sigma=0$), where the islands all disappear and the strips become
extended much further on the downward diagonal. Inspecting the fits
individually indicates that mean squared deviation does not give an
adequate measure of fit quality in this special case. This anomaly seems
due to the fact that, in the absence of Gaussian noise from experimental
statistics, systematic deviations from (\ref{jlogM}) associated with the
approximation (\ref{prodT}) become important.

\section{Summary procedure for mechanism identification from control
experimental data}\label{summary}

In order to extract the basic mechanism information comprising
$j_\mathrm{min}$ and $\langle j_\mathcal{P} \rangle$ from quantum control
experimental data, the methods of \S\ref{simulated} can be distilled into
the following general procedure:

\begin{enumerate}

\item Perform a closed-loop optimization of population transfer, giving an
associated optimal laser pulse shape $E(t)$.

\item Apply a modulated field $\tilde E(t) = \mcM E(t)$ to the system and
measure the the resulting final state population $|\tilde
\psi_{n_\mrf}(t_\mrf)|^2$ for many values of $\mcM$ over some range
$\sim(\delta,1.5)$, where $\delta$ is a positive value near 0 determined
by experimental sensitivity.

\item Extract $j_\mathrm{min}$ from the data by extrapolating the limit
(\ref{dlogpsidlogM}).

\item Choose a truncation of the sum in (\ref{jlogM}), \emph{e.g.}\
$k_\mathrm{max}=4$, and perform a non-linear fit to the data for each of a
set of fitting ranges $(\mcM_\mathrm{min},\mcM_\mathrm{max})$,
\emph{e.g.}\ the set (\ref{range}). One may choose to constrain the fit by
requiring $a > j_\mathrm{min}$ in (\ref{jlogM}).

\item Plot the fit values of $\langle j_\mathcal{P} \rangle$, as in Fig.\
\ref{javg}, and the corresponding mean squared deviations, as in Fig.\
\ref{fq}, over the $\mcM_\mathrm{min}$--$\mcM_\mathrm{max}$ plane. Exclude
regions in which the fit violates the condition $\langle j_\mathcal{P}
\rangle > j_\mathrm{min}$ and pathological regions like in the lower-left
corner of Fig.\ \ref{fq}.

\item Find the point $(\mcM_\mathrm{min},\mcM_\mathrm{max})$ at which the
mean squared deviation is minimized, giving the associated fit value of
$\langle j_\mathcal{P} \rangle$ as that most consistent with the data.

\end{enumerate}

\section{Remarks and Conclusion}

Since Bell's model can be defined for any choice of basis $|n\rangle$,
there is a more general question of how mechanism analysis varies with the
choice of basis. Beyond that, Bell's jump rule (\ref{T}) itself permits
generalization \cite{Guido}, providing additional freedom over which
trajectory probability assignments may vary. The import of this freedom
for mechanism identification remains to be determined.

This paper has shown how Bell's beable model of quantum mechanics can be
used to understand the dynamics of quantum systems driven by complicated
optimal control fields. Beable trajectories are identified with simple
physical processes effecting the controlled transfer of population from
one state to another, and aggregations of beable trajectories may be used
to compute the importance of different such processes in the dynamics. In
the context of a model 7-level system, numerical simulations reveal four
chief pathways and also a host of higher order pathways that are
collectively significant on the 40\% population level. We have shown how
the control field sweeps trajectories into these pathways by switching on
and off beable flow over specific transitions on a fs time-scale.

Beable trajectory methods were then defined in general for extracting
statistical mechanism information directly from experimental data, without
requiring knowledge of a Hamiltonian or even the level structure of the
system under study. Application to simulated noisy data for the model
system produced the correct minimum number of quantum transitions in the
control process and the average number of such transitions to within 3\%
at noise levels up to 25\%.

\section*{Acknowledgement}

The authors acknowledge support from the NSF and DoD. ED acknowledges
partial support from the Program for Plasma Science and Technology at the
Princeton Plasma Physics Laboratory.

\pagebreak

\begin{table}
\begin{center}
\begin{tabular}{|l|l|}
\hline
probability & pathway \\
\hline
0.19  & 0 2 3 5 6 \\
0.16  & 0 2 3 4 6 \\
0.14  & 0 1 3 5 6 \\
0.12  & 0 1 3 4 6 \\
0.018  & 0 2 3 5 6 5 6 \\
\hline
0.005  & 0 2 \\
\hline
0.0007  & 0 2 3 5 6 4 3 5 6 \\
\hline
\end{tabular}
\caption{\small The five most probable pathways, followed by the highest
probability pathway failing to reach $n=6$ at $t=100$ fs, and then the
highest probability pathway involving a topologically non-trivial cycle
in state space. The fractional error in the pathway probability $P$ is
given roughly by $(10^5 P)^{-1/2}$.}
\label{tab}
\end{center}
\end{table}

\begin{thebibliography}{}

\bibitem{control} R. Judson, H. Rabitz, Phys. Rev. Lett. \textbf{68}, 1500
(1992).

\bibitem{organic} R. Levis, G. Menkir, H. Rabitz, Science \textbf{292},
709 (2001).

\bibitem{metal} A. Assion et al., Science, \textbf{282}, 919 (1998).

\bibitem{Na2K} S. Vajda et al., Chem. Phys., \textbf{267}, 231 (2001).

\bibitem{AlGaAs} J. Kunde et al., Appl. Phys. Lett. \textbf{77}, 924
(2000).

\bibitem{Xray} R. Bartels et al., Nature \textbf{406}, 164 (2000).

\bibitem{dye} C. Bardeen et al., Chem. Phys. Lett. \textbf{280}, 151
(1997).

\bibitem{CH3OH} T. Weinacht, J. White, P. Bucksbaum, J. Chem. A
\textbf{103}, 10166 (1999). 

\bibitem{Dahleh} M. Dahleh, A. Peirce, H. Rabitz, Phys. Rev. A
\textbf{37}, 4950 (1988).

\bibitem{Kosloff} R. Kosloff et al., Chem. Phys. \textbf{139}, 201 (1989).

\bibitem{Zhu} H. Rabitz, W. Zhu, Accts. Chem. Res. \textbf{33}, 572
(2000).

\bibitem{Rice} S. Rice, M. Zhao, \emph{Optical Control of Molecular
Dynamics}, Wiley (2000).

\bibitem{Mitra1} An alternative scheme quantifies pathway importance by
associating to each pathway not a probability but rather an amplitude and
a phase (A. Mitra and H. Rabitz, to be published). Although this latter
scheme was not originally formulated in terms of dynamical trajectories,
the amplitudes correspond in some sense to the trajectory probabilities
that result from the jump rule $T_{nm}=|H_{nm}|/\hbar$, which does not
preserve (\ref{modpsi}).

\bibitem{Mitra2} Optimization results provided by A. Mitra.

\bibitem{Bohm} D. Bohm, Phys. Rev. \textbf{85}, 166 (1952); Phys. Rev.
\textbf{85}, 180 (1952).

\bibitem{Bell} J. Bell, ``Beables for quantum field theory,''
\emph{Speakable and unspeakable in quantum mechanics}, Cambridge
University Press (1987).

\bibitem{intpic} The definition of $|n\rangle$ as interaction picture
states has the affect of eliminating larger contributions to the jump
probabilities $T_{nm}$ from the $\hbar \omega$ terms, hence reducing the
overall frequency of jumps. Had we taken $|n\rangle$ as Schrodinger
picture states, we would have had to decrease the time step by a factor
$\hbar \omega/\mu E$ for comparable results. This factor is around 10 for
the model 7-level system.

\bibitem{Guido} G. Bacciagaluppi, Found. Phys. Lett. \textbf{12} 1 (1999),
quant-ph/9811040.

\end{thebibliography}
\end{document}